\documentclass[journal]{IEEEtran} 

\usepackage{ifthen}
\newboolean{DoubleColumn}
\setboolean{DoubleColumn}{true} 

\usepackage[font=footnotesize]{caption}

\usepackage{balance}
\usepackage{mathtools}
\usepackage{amsmath,amsfonts,amssymb}
\usepackage{scalerel}
\usepackage[utf8]{inputenc}
\usepackage{graphicx}
\usepackage{epstopdf}
\usepackage{ellipsis}
\usepackage{algorithmic}
\usepackage{algorithm}
\usepackage{physics}
\usepackage{svg}
\usepackage{lipsum}
\usepackage{siunitx}
\ifCLASSOPTIONcompsoc
 \usepackage[caption=false,font=normalsize,labelfont=sf,textfont=sf]{subfig}
\else
 \usepackage[caption=false,font=footnotesize]{subfig}
\fi
\usepackage[nolist]{acronym}
\usepackage[noadjust]{cite}
\usepackage{academicons}
\usepackage{tikz}
\usetikzlibrary{patterns,positioning,intersections}
\usetikzlibrary{arrows}
\usetikzlibrary{svg.path}
\usetikzlibrary{patterns}
\usetikzlibrary{calc}
\usetikzlibrary{positioning}
\usetikzlibrary{decorations.pathreplacing,decorations.markings,shapes.geometric}

\begin{acronym}
\acro{PCL}{passive coherent location}
\acro{3D}{three-dimensional}
\acro{RCS}{radar cross-section}
\acro{DFL}{device-free localization}
\acro{MDFL}{multipath-enhanced device-free localization}
\acro{FIM}{Fisher information matrix}
\acro{LoS}{line-of-sight}
\acro{GLoS}{geometrical line-of-sight}
\acro{GPS}{global positioning system}
\acro{KEST}{Kalman enhanced super resolution tracking}
\acro{EKF}{extended Kalman filter}
\acro{KF}{Kalman filter}
\acro{CIR}{channel impulse response}
\acro{AoA}{angle of arrival}
\acro{TDoAoMC}{time difference of arrival between multipath components}
\acro{FCC}{Federal Communications Commission}
\acro{E-911}{enhanced 911}
\acro{GNSS}{global navigation satellite system}
\acro{NLoS}{non-line-of-sight}
\acro{UWB}{ultra-wideband}
\acro{WLAN}{wireless local area network}
\acro{VT}{virtual transmitter}
\acro{VR}{virtual receiver}
\acro{VN}{virtual node}
\acro{RP}{reflection point}
\acro{MPC}{multipath component}
\acro{TDOA}{time difference of arrival}
\acro{TOA}{time of arrival}
\acro{PDF}{probability density function}
\acro{SLAM}{simultaneous localization and mapping}
\acro{SNR}{signal-to-noise-ratio}
\acro{PF}{particle filter}
\acro{SAGE}{space-alternating generalized expectation-maximization}
\acro{MMSE}{minimum mean square error}
\acro{RMSE}{root mean square error}
\acro{MSE}{mean square error}
\acro{CDF}{cumulative distribution function}
\acro{DOP}{dilution of precision}
\acro{CRLB}{Cram\'{e}r-Rao lower bound}
\acro{PCRLB}{posterior Cram\'{e}r-Rao lower bound}
\acro{ISI}{intersymbol interference}
\acro{RBPF}{Rao-Blackwellized particle filter}
\acro{SoO}{signals of opportunity}
\acro{SISO}{single-input single-output}
\acro{RF}{radio frequency}
\acro{IMU}{inertial measurement unit}
\acro{SIS}{sequential importance sampling}
\acro{SISPF}{sequential importance sampling particle filter}
\acro{SIR}{sequential importance resampling}
\acro{SIRPF}{sequential importance resampling particle filter}
\acro{3GPP-LTE}{3rd generation partnership project - long-term evolution}
\acro{TDoAbMC}{time difference of arrival between multipath components}
\acro{subPF}{subordinate particle filter}
\acro{superPF}{superordinate particle filter}
\acro{DLL}{delay locked loop}
\acro{RFID}{radio  frequency  identification}
\acro{RSS}{received signal strength}
\acro{AWGN}{additive white Gaussian noise}
\acro{MUSIC}{multiple signal classification}
\acro{EM}{expectation maximization}
\acro{ML}{maximum likelihood}
\acro{PSD}{power spectral density}
\acro{SRA}{super resolution algorithm}
\acro{MAP}{maximum a posteriori}
\acro{LS}{least square}
\acro{DBN}{dynamic Bayesian network}
\acro{MEMS}{micro-electro-mechanical system}
\acro{INS}{inertial navigation systems}
\acro{OFDM}{orthogonal frequency-division multiplexing}
\acro{HMM}{hidden Markov model}
\acro{ZUPT}{zero velocity updates}
\acro{DMC}{Dense Multipath Components}
\acro{TS}{track-section}
\acro{log-domain}{logarithmic domain}
\acro{lin-domain}{linear domain}
\acro{bframe}[b-frame]{body frame}
\acro{iframe}[i-frame]{ineratial frame}
\acro{nframe}[n-frame]{navigation frame}
\acro{logCDF}[log-CDF]{cumulative distribution function in \ac{log-domain}}
\acro{logweight}[log-weight]{log-domain weight}
\acro{Lin-PF}[Lin-PF]{linear domain PF}
\acro{Log-PF}[Log-PF]{logarithmic domain PF}
\acro{LogMAP}[Log-PF MAP]{\ac{MAP} of Log-PF}
\acro{LogMSE}[Log-PF \acs{MMSE}]{\ac{MMSE} of Log-PF}
\acro{LogMAX}[LogPF MAX]{\ac{MAX} of LogPF}
\acro{LinMAP}[PF MAP]{\ac{MAP} of linear domain \ac{PF}}
\acro{LinMSE}[PF \acs{MMSE}]{\ac{MMSE} of linear domain \ac{PF}}
\acro{LinMAX}[PF MAX]{\ac{MAX} of linear domain \ac{PF}}
\acro{Receiver1}[receiver 1]{receiver 1}
\acro{Receiver2}[receiver 2]{receiver 2}
\acro{Receiver3}[receiver 3]{receiver 3}
\acro{MovIMU}[IMU-Transition-Model]{IMU-Transition-Model}
\acro{MovNoIMU}[Gaussian-Transition-Model]{Gaussian-Transition-Model}
\acro{algoChannelSLAMNoIMU}[NoIMU-Channel-SLAM]{NoIMU-Channel-SLAM}
\acro{RedChannelSLAM}[Dynamic-Channel-SLAM]{Dynamic-Channel-SLAM}
\acro{RBPFChannelSLAM}[RBPF-Channel-SLAM]{RBPF-Channel-SLAM}
\acro{LPF}{likelihood particle filter}
\end{acronym} 

\acresetall 

\definecolor{orcidlogocol}{HTML}{A6CE39}
\tikzset{
  orcidlogo/.pic={
    \fill[orcidlogocol] svg{M256,128c0,70.7-57.3,128-128,128C57.3,256,0,198.7,0,128C0,57.3,57.3,0,128,0C198.7,0,256,57.3,256,128z};
    \fill[white] svg{M86.3,186.2H70.9V79.1h15.4v48.4V186.2z}
                 svg{M108.9,79.1h41.6c39.6,0,57,28.3,57,53.6c0,27.5-21.5,53.6-56.8,53.6h-41.8V79.1z M124.3,172.4h24.5c34.9,0,42.9-26.5,42.9-39.7c0-21.5-13.7-39.7-43.7-39.7h-23.7V172.4z}
                 svg{M88.7,56.8c0,5.5-4.5,10.1-10.1,10.1c-5.6,0-10.1-4.6-10.1-10.1c0-5.6,4.5-10.1,10.1-10.1C84.2,46.7,88.7,51.3,88.7,56.8z};
  }
}

\newcommand\orcidicon[1]{\href{https://orcid.org/#1}{~\mbox{\scalerel*{
\begin{tikzpicture}[yscale=-1,transform shape]
\pic{orcidlogo};
\end{tikzpicture}
}{|}}}}

\usepackage[hidelinks]{hyperref}

\title{Multipath-Enhanced Device-Free Localization\\ in Wideband Wireless Networks}
\author{
\IEEEauthorblockN{Martin~Schmidhammer\textsuperscript{\orcidicon{0000-0002-9345-142X}},~\IEEEmembership{Member,~IEEE},
Christian~Gentner\textsuperscript{\orcidicon{0000-0003-4298-8195}},~\IEEEmembership{Member,~IEEE},
Stephan~Sand\textsuperscript{\orcidicon{0000-0001-9502-5654}},~\IEEEmembership{Senior~Member,~IEEE},
Uwe-Carsten~Fiebig\textsuperscript{\orcidicon{0000-0003-2736-1140}},~\IEEEmembership{Member,~IEEE}\\}
\thanks{The authors are with the Institute of Communications and Navigation, German Aerospace Center (DLR), 82234 Wessling, Germany (e-mail: \mbox{martin.schmidhammer@dlr.de;} christian.gentner@dlr.de; stephan.sand@dlr.de; uwe.fiebig@dlr.de).}
}

\renewcommand{\baselinestretch}{0.97}

\begin{document}
\maketitle
\IEEEoverridecommandlockouts
\IEEEpubid{\scriptsize This work has been submitted to the IEEE for possible publication. Copyright may be transferred without notice, after which this version may no longer be accessible.}

\begin{abstract}
State-of-the-art \acl{DFL} systems infer presence and location of users based on \acl{RSS} measurements of \acl{LoS} links in wireless networks.
In this letter, we propose to enhance \acl{DFL} systems by exploiting multipath propagation between the individual network nodes.
Particularly indoors, wireless propagation channels are characterized by multipath propagation, i.e., received signals comprise \aclp{MPC} due to reflection and scattering.
Given prior information about the surrounding environment, e.g., a floor plan, the individual propagation paths of \aclp{MPC} can be derived geometrically.
Inherently, these propagation paths differ spatially from the \acl{LoS} propagation path and can be considered as additional links in the wireless network.
This extended network determines the novel multipath-enhanced \acl{DFL} system. 
Using theoretical performance bounds on the localization error, we show that including \aclp{MPC} into \acl{DFL} systems improves the overall localization performance and extends the effective observation area significantly.
\end{abstract}

\begin{IEEEkeywords}
device-free localization, wireless sensor networks, multipath propagation, Cramér-Rao lower bounds.
\end{IEEEkeywords}


\section{Introduction}
Ubiquitous connectivity and location-based services are key components for smart environments, such as modern manufacturing facilities and smart homes~\cite{shit2019}.
This demand in location awareness can be served e.g., by active \ac{RF}-based localization systems requiring the user to carry a localization device.
Alternatively, novel passive localization systems estimate the location of the user through measuring the user’s effects on \ac{RF} signals.
The user does not need to carry any communication device, thus, we speak about \ac{DFL}~\cite{patwari2010}.
%
Exploiting user induced fading effects, such as diffraction and shadowing, \ac{DFL} systems measure typically the \ac{RSS} values between network nodes along the \ac{LoS} paths to infer presence and location of the user.
Based on these \ac{RSS} measurements, the location is either estimated by computing propagation field images, so-called radio tomographic imaging~\cite{wilson2010}, or by using empirical~\cite{guo2015} or theoretical propagation models~\cite{rampa2015}, which directly relate the \ac{RSS} measurements to the user location.
%
%
%
\IEEEpubidadjcol

Prevalent \ac{DFL} systems deploy narrowband \ac{RF} devices, and thus, besides the user impact, the \ac{RSS} is also affected by small scale multipath fading.
Since multipath fading is unique for each network link, narrowband \ac{DFL} systems require extensive initial calibration and frequent recalibration accounting for time-variant propagation environments~\cite{kaltiokallio2017_arti}.
%
Addressing the issues of multipath fading, the authors of~\cite{beck2016} propose to use \acl{UWB} \ac{RF} devices for \ac{DFL} systems.
Thereby, the large signal bandwidth allows to mitigate the effects due to multipath propagation by separating the \ac{LoS} path from reflected paths.
The user induced fading can be thus determined deterministically, which enhances the performance and reduces the calibration efforts of the \ac{DFL} system.

\IEEEpubidadjcol

\begin{figure}
	\centering
	\includegraphics[width=0.95\columnwidth]{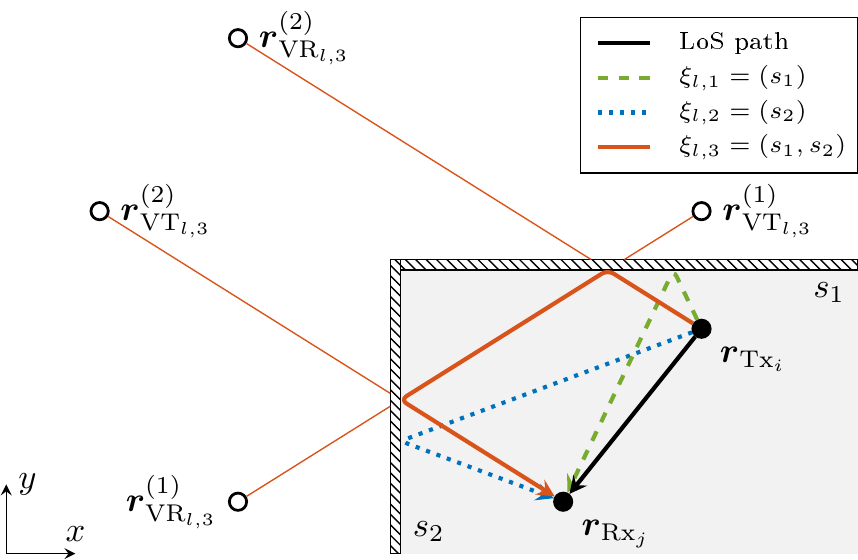}
	\setlength{\belowcaptionskip}{-10pt}
	\caption{\footnotesize Exemplary multipath propagation for network link $l$ $\left(\mathrm{Tx}_i\right.$~and~$\left.\mathrm{Rx}_j\right)$ in a given environment with two reflecting surfaces $\mathcal{S}=\{s_1,s_2\}$. 
%
	Arrows indicate the physical propagation paths of \acs{LoS} and \acsp{MPC} according to the set of visible sequences
	$\mathcal{X}_l = \{\xi_{l,1},\xi_{l,2},\xi_{l,3}\}$.
	The geometric decomposition is illustrated for the second-order reflection $\xi_{l,3}$ by virtual transmitters and receivers at mirrored positions of $\mathrm{Tx}_i$ and $\mathrm{Rx}_j$.
	Resulting equidistant paths between pairs of corresponding nodes are indicated by red lines reconstructing the physical propagation path of the \acs{MPC} within the observation area.
}
\label{fig:geometry}
\end{figure}
Based on wideband signal measurements, we have demonstrated in~\cite{schmidhammer2020} that user induced fading can also be observed in the received power of reflected and scattered signals.
Naturally, the propagation paths of reflected and scattered signals differ spatially from the \ac{LoS} propagation path (see Fig.~\ref{fig:geometry}), and thus, contain spatial information which can be used in addition to the spatial information embedded in \ac{LoS} paths.
Instead of mitigating, we therefore propose to make particularly use of multipath propagation for \ac{DFL}.
The goal of this letter is to illustrate the performance improvement of \ac{DFL} systems that can be obtained by incorporating propagation paths of reflected and scattered signals as additional links to the underlying wireless network.
We provide the signal processing to extract the spatial information from each pair of transmitting and receiving node and derive the theoretical performance bounds on the localization error for the novel \ac{MDFL} system.
In a case study, we evaluate \ac{MDFL} numerically and quantify the improvement compared to common \ac{DFL}.


\section{Network and Propagation Model}
\label{sec:system}
We consider a \ac{MDFL} system relying on a network of $N_{\mathrm{Tx}}$~transmitting and $N_{\mathrm{Rx}}$ receiving nodes at known locations
${\boldsymbol{r}_{\mathrm{Tx}_i}}$, ${i\in \{1,\dots,N_{\mathrm{Tx}}\}}$, and
${\boldsymbol{r}_{\mathrm{Rx}_j}}$, ${j\in \{1,\dots,N_{\mathrm{Rx}}\}}$.
Receiving nodes can be collocated with transmitting nodes or individually placed.
The network link configuration is determined by the index set $\mathcal{P}$, where link ${(i,j) \in \mathcal{P}}$ is composed of the $i$-th transmitting and the $j$-th receiving node and is indexed by $l\in\{1,\dots,\left|\mathcal{P}\right|\}$.
For link $l$, the signal at the receiving node is modeled as a superposition of scaled and delayed replica of a known transmit signal $s_l(t)$ of duration $T_{\mathrm{sym}}$.
These comprise the \ac{LoS} and a finite number of 
static \acp{MPC} due to reflections off the surrounding environment.
Therewith, the received signal is expressed as
\begin{equation}
\label{eq:signal_model}
y_{l}(t) = \sum_{m=1}^{N_{l}} \alpha_{l,m} s_l(t-\tau_{l,m}) + n_{l}(t),
\end{equation}
with time-variant, complex amplitude $\alpha_{l,m}$ and static propagation delay $\tau_{l,m}$ of the $m$-th \ac{MPC}~\cite{molisch2009}. 
For notational convenience, we omit a time index for the amplitude and consider \ac{LoS} paths also as \acp{MPC}.
The term $n_{l}(t)$ denotes white circular symmetric normal distributed noise with variance $\sigma^2_{y_l}$.

Following \cite{meissner2014} and \cite{gentner2016}, we model the delays of the 
\acp{MPC} geometrically using virtual nodes. 
Therefore, we represent the surrounding environment by a finite number of reflecting surfaces determining the set $\mathcal{S}$.
{\color{black}
For this representation we require prior information about the surrounding envirionment, which can be provided, e.g., by a floor plan~\cite{meissner2014}.
Based on the set of reflecting surfaces $\mathcal{S}$, we can define reflection sequences that chronologically describe the signal propagation for an \ac{MPC} $p$ from the transmitting to the receiving node of link~$l$.}
{\color{black}
Using tuple notation we express these sequences as 
${\xi_{l,p} = (s_{b})}$
with $s_{b}\in\mathcal{S}$,
where the sequence length, denoted as~$N_{\xi_{l,p}}$, is determined by the order of reflection.
%
Subsequently, we can compose a set $\mathcal{X}$ containing all potential reflection sequences.
}
Based on these sequences, we can construct \acp{VT} and \acp{VR} for each \ac{MPC} of the network links by consecutively mirroring the physical nodes.
Note that due to symmetry, \acp{VR} are constructed using the sequences in reverse order.
For sequence $\xi_{l,p} \in \mathcal{X}$, the locations of the virtual nodes are thus denoted by
$\boldsymbol{r}_{\mathrm{VT}_{l,p}}^{(u)}$ and
$\boldsymbol{r}_{\mathrm{VR}_{l,p}}^{(N_{\xi_{l,p}}-u)}$,
{\color{black}
where the index
${u\in \{0,\dots,N_{\xi_{l,p}}\}}$
corresponds to pairs of related virtual nodes.
}
Thereby, the physical transmitting and receiving nodes are referred to as
$\boldsymbol{r}_{\mathrm{VT}_{l,p}}^{(0)}$ and 
$\boldsymbol{r}_{\mathrm{VR}_{l,p}}^{(0)}$, respectively.
Fig.~\ref{fig:geometry} provides an example for a second-order reflection illustrating the sets of \acp{VT} and \acp{VR}.
{\color{black}
It is noticeable that the distances between the pairs of related nodes are equal.
Due to geometry these distances correspond to the length of the physical propagation path.
Thus, we can express this length for any pair of related nodes $u$ as
}
\begin{equation}
\label{eq:distance}
d\left(\xi_{l,p}\right) = d_{l,p} = \lVert \boldsymbol{r}_{\mathrm{VT}_{l,p}}^{(u)}-\boldsymbol{r}_{\mathrm{VR}_{l,p}}^{(N_{\xi_{l,p}}-u)}\rVert.
\end{equation}
%
Furthermore, as shown in Fig.~\ref{fig:geometry}, the paths between pairs of related nodes intersect at the physical \aclp{RP}.
Thus, we can reconstruct the physical propagation paths geometrically similar to optical ray-tracing~\cite{meissner2014}.
This allows to compose a set of visible sequences~$\mathcal{X}_l \subseteq \mathcal{X}$ for each network link, with cardinality $|\mathcal{X}_l| = N_l$ equal to the number of \acp{MPC} modeled in~(\ref{eq:signal_model}).
Finally, we can define a set of expected path lengths as
\begin{equation}
\label{eq:set_expPL}
\mathcal{D}_l = \{d({\xi}_{l,m})\,|\,{\xi}_{l,m}\in \mathcal{X}_l \}.
\end{equation}

\section{Multipath-enhanced Device-Free Localization}
\label{sec:mdfl}
The objective of any \ac{DFL} system
is to estimate the user state, defined as location and velocity, based on user-induced changes in the received signal power.
Following a Bayesian approach, this objective can be expressed by a transition model describing the spatio-temporal evolution of the user state, and a measurement model relating the measured changes in the received signal power to the user state.
In the following, we describe the required signal processing and provide a corresponding measurement model.


{\color{black}\subsection{Initialization and Data Association}\label{sec:mdfl_ini}}
Initially, we need to determine the individual propagation effects of the static environment.
Therefore, the channel of each network link is observed over an initialization period~$T_{\mathrm{ini}}$.
{\color{black}
Ideally, the environment should be devoid of any user during this period.
This ensures that we can accurately describe the propagation effects of the static environment.
}%
With~$T_{\mathrm{g}}$ as time interval between two adjacent received signals, a total of~%
${\lfloor T_{\mathrm{ini}}/T_{\mathrm{g}} \rfloor}$ 
consecutive signal samples are collected.
{\color{black}
For each signal sample, we determine amplitude and delay values for $\hat{N}_l$ seperable \acp{MPC} using \acl{ML} estimation,
}
%
e.g., using the \ac{SAGE} algorithm~\cite{fleury1999}.

{\color{black}
By averaging the amplitude and delay estimates over the amount of signal samples, we obtain 
the set of mean amplitude
$\{\bar{\alpha}_{l,q}\}_{q=1}^{\hat{N}_l}$
and
the set of mean delay
$\{\bar{\tau}_{l,q}\}_{q=1}^{\hat{N}_l}$.
The set of mean amplitude allows to calculate the power of the \acp{MPC} for the idle channel, which serves as reference to determine user induced power changes.
The set of mean delay determines the set of estimated path lengths as
}
\begin{equation}
\label{eq:set_estPL}
\hat{\mathcal{D}}_l = \{c \cdot \bar{\tau}_{l,q} \,|\, 1 \leq q \leq \hat{N}_l\}.
\end{equation}

\begin{figure*}[t!]
	\centering
	\ifthenelse {\boolean{DoubleColumn}}%
	{%
	\centering
   	\subfloat[DFL system measuring LoS only.\label{fig:crlb_los}]{%
	   	\includegraphics[width=0.95\columnwidth]{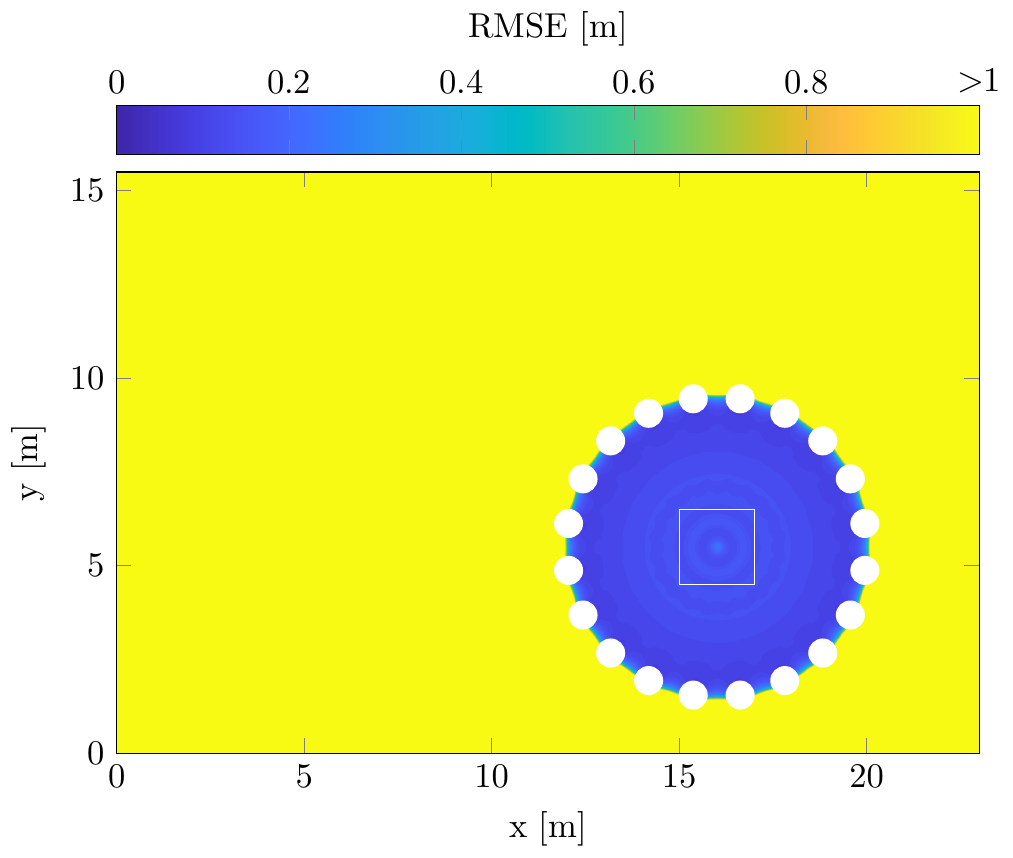}}
	\centering
	\hspace*{0.1\columnwidth}
   	\subfloat[MDFL system measuring both LoS and MPCs. \label{fig:crlb_mpc}]{%
	   	\includegraphics[width=0.95\columnwidth]{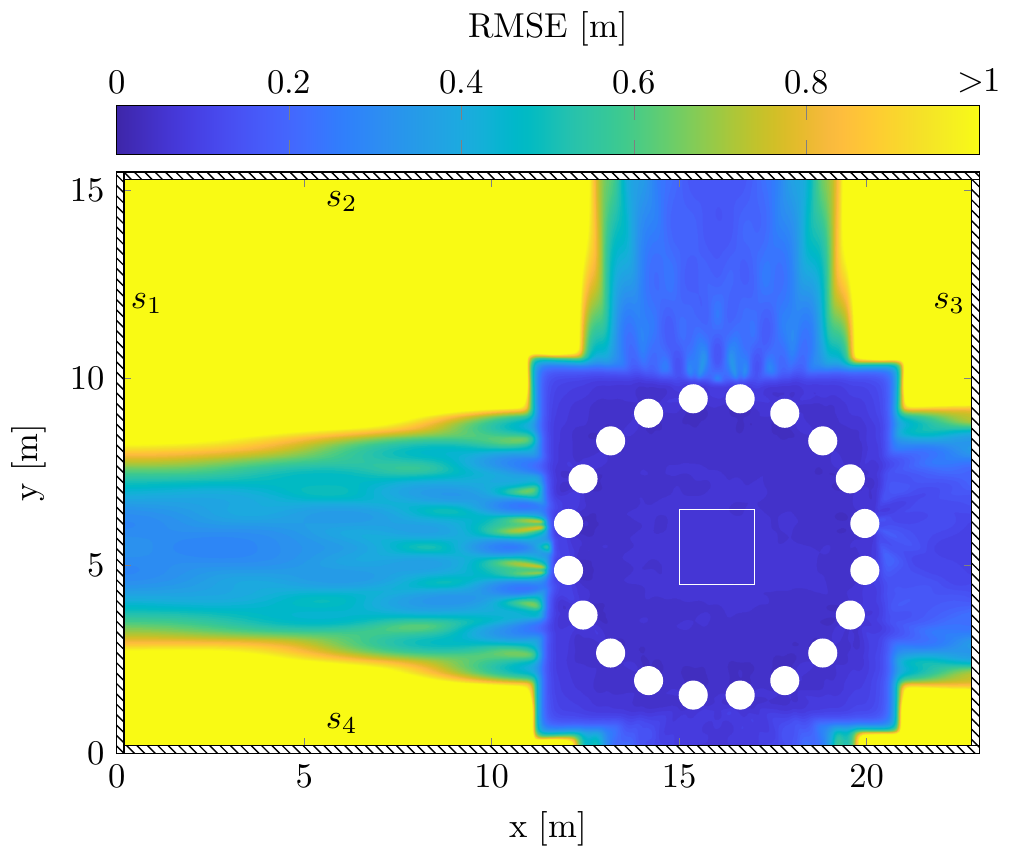}}
	}%
	{%
   	\subfloat[DFL system measuring LoS only.\label{fig:crlb_los}]{%
	   	\includegraphics[width=0.5\columnwidth]{figures/crlb_mdfl/circle_no_wall.pdf}}
   	\subfloat[MDFL system measuring both LoS and MPCs. \label{fig:crlb_mpc}]{%
	   	\includegraphics[width=0.5\columnwidth]{figures/crlb_mdfl/circle_4_wall.pdf}}
	}%
	\setlength{\belowcaptionskip}{-10pt}
  	\caption{\footnotesize CRLB limit to the localization accuracy of (a) DFL and (b) MDFL. 
  	The underlying network is identical for both systems, composed of \SI{20}{} circular arranged transceiving nodes as indicated by white bold dots.
  	In (b), the reflecting surfaces $\mathcal{S}=\{s_1,s_2,s_3,s_4\}$ are represented by hatched lines.
  	The illustrated \ac{MDFL} system considers \acp{MPC} exemplarily only due to first-order reflections.
	The white rectangle in the network center defines an area of $\SI{2}{\m} \times \SI{2}{\m}$ used to calculate the expected RMSE (cf.~Fig.~\ref{fig:crlb_numDevices}).
  }
  \label{fig:crlb} 
\end{figure*}

{\color{black}
After the initialization, i.e., the estimation of all observable \acp{MPC} from the received signals, we need to determine the physical propagation paths of these \acp{MPC}.
In Sec.~\ref{sec:system} we have modeled the delays of \acp{MPC} using virtual nodes and reflection sequences.
Thereby, a reflection sequence describes chronologically the signal propagation from transmitting to receiving node and is thus a representation of the physical propagation path.
Therefore, we can relate the \acp{MPC} estimated during initialization to the physical propagation paths by associating these \acp{MPC} with the modeled \acp{MPC} and their corresponding reflection sequences.




%
}
{\color{black}
The \acp{MPC} are characterized by delay or path length, respectively.
Therefore, we use the corresponding sets of expected and estimated path lengths for data association, i.e., $\mathcal{D}_l$ in (\ref{eq:set_expPL}) and $\hat{\mathcal{D}}_l$ in (\ref{eq:set_estPL}).
}
A possible association approach provides~\cite{meissner2014}, following optimal subpattern assignment~\cite{schuhmacher2008}.
Thereby, the sets $\mathcal{D}_l$ and $\hat{\mathcal{D}}_l$ are matched such that the cumulative distance between expected and estimated path lengths is minimized.
The individual associations are further constrained not to exceed a certain distance value, the so-called cut-off value.
{\color{black}
The cut-off value is defined by the ranging accuracy of the transmitted signals~\cite{meissner2014}.
Therewith, we discard strongly outlying \acp{MPC} and avoid wrong associations.
Clutter due to diffuse reflections and scattering is thus inherently eliminated.}

{\color{black}
This constrained data association approach is applied for each network link $l$ resulting in sets of associated reflection sequences~$\mathcal{X}_l^{\star}$.
Depending on the observed \acp{MPC} during initialization and the subsequent data association, the set~$\mathcal{X}_l^{\star}$ of network link $l$ contains an individual number of $N_l^{\star}$ associated sequences.
Combining the information from all network links results in 
}
the union set~${\mathcal{X}^{\star} = \cup_{l=1}^{|\mathcal{P}|}\mathcal{X}_l^{\star}}$. 
{\color{black}
Eventually, the cardinality
$|\mathcal{X}^{\star}| = \sum_{l=1}^{|\mathcal{P}|}N_l^*$
determines the overall amount of associated reflection sequences and thus the amount of corresponding \acp{MPC} which can be used for \ac{MDFL}.
}

{\color{black}
\subsection{Parameter Estimation and Measurement Model}
\label{sec:mdfl_est}}
After initialization
{\color{black} and data association, the network is basically ready for localization,
but requires measurement data.
Therefore we continuously determine the amplitude values of all associated \acp{MPC} of each network link.
}
For each $\xi_{l,n}\in\mathcal{X}^{\star}$ with corresponding delay $\bar{\tau}_{l,n}$
{\color{black}
we estimate the amplitude using
}
\begin{equation}
\label{eq:amplitude_est}
\hat{\alpha}_{l,n} = \hat{\alpha}(\bar{\tau}_{l,n}) = \int_0^{T_{sym}} y_{l,n}^{\mathrm{res}}(t)^{*} s_l(t-\bar{\tau}_{l,n}) \, d t
\end{equation}
as the projection of the residuum signal 
$y_{l,n}^{\mathrm{res}}(t)$
onto the unit transmit signal $s_l(t)$~\cite{meissner2014}.
Thereby, the residuum signal is defined as received signal adjusted for all \acp{MPC} up to the $(n-1)$-th, i.e.,
$y_{l,n}^{\mathrm{res}}(t) =y_{l}(t)- \sum_{n'=1}^{n-1}\hat{\alpha}_{l,n'}s_l(t-\bar{\tau}_{l,n'})$.

Given the amplitude estimates, we can calculate the measured power of an \ac{MPC} as $|\hat{\alpha}_{l,n}|^2$ and express the user induced power changes by 
adjusting the measured power by the power of the idle channel determined during initialization.
Thus, we can compose the measurement vector 
${\boldsymbol{z} \in \mathbb{R}^{|\mathcal{X}^{\star}|\times 1}}$
by stacking the individual components 
$z_{l,n}=20\log_{10}\frac{|\hat{\alpha}_{l,n}|}{|\bar{\alpha}_{l,n}|}$.

%
In the follwing, we model the measurement vector as
\begin{equation}
\label{eq:measurement_model}
\boldsymbol{z} = \left[ \dots,h(\boldsymbol{r},\xi_{l,n}),\dots\right]^{\mathrm{T}} + \boldsymbol{w}, \quad \forall l,n : \xi_{l,n}\in \mathcal{X}^{\star},
\end{equation}
with Gaussian measurement noise $\boldsymbol{w} \sim \mathcal{N}(\boldsymbol{0},\boldsymbol{R})$. 
Assuming mutually independent measurements, the noise covariance matrix $\boldsymbol{R}\in \mathbb{R}^{|\mathcal{X}^{\star}|\times|\mathcal{X}^{\star}|}$ is diagonal and defined as 
\begin{equation}
\label{eq:noise_cov}
\boldsymbol{R} = \mathrm{diag}(\dots,\sigma_{l,n}^2,\dots), \quad  \forall l,n : \xi_{l,n}\in \mathcal{X}^{\star}.
\end{equation}
%
As shown in~\cite{schmidhammer2020}, user-induced variations in the {\color{black} power} of an \ac{MPC} can be modeled by superimposing the user impact on each pair of related virtual nodes.
In this work, the impact on the individual node pairs is calculated using the empirical exponential model~\cite{guo2015}.
Depending on the user location $\boldsymbol{r}$,
{\color{black}
which is defined as the body center of the user,
}
we can thus express changes in the {\color{black} power} of the \ac{MPC} modeled by sequence $\xi_{l,n}$ as the sum
\begin{equation}
\label{eq:exp_model_sumMPC}
h(\boldsymbol{r},\xi_{l,n}) = \sum_{u=0}^{N_{\xi_{l,n}}} \phi_{l,n} e^{-\delta_{l,n}^{(u)}(\boldsymbol{r})/\kappa_{l,n}},
\end{equation}
with the parameter $\phi_{l,n}$ and $\kappa_{l,n}$ defining the maximum modeled change in {\color{black} power} and the decay rate.
The excess path length $\delta_{l,n}^{(u)}(\boldsymbol{r})$ corresponds to the \mbox{$u$-th} pair of virtual nodes as defined by sequence $\xi_{l,n}$ and is calculated by
\begin{equation}
\delta_{l,n}^{(u)}(\boldsymbol{r}) =
\Vert \boldsymbol{r}_{\mathrm{VT}_{l,n}}^{(u)} - \boldsymbol{r} \Vert + \Vert \boldsymbol{r}_{\mathrm{VR}_{l,n}}^{(N_{\xi_{l,n}}-u)} - \boldsymbol{r} \Vert - d_{l,n},
\end{equation}
with $d_{l,n}$ as path length defined in~(\ref{eq:distance}).

\section{Performance Bound}
\label{sec:CRLB}
The \ac{CRLB} provides a lower bound on the variance of an unbiased estimator, defined by the inverse of the \ac{FIM}.
Therefore, the unbiased estimator~$\hat{\boldsymbol{r}}$ of user location 
{\color{black}${\boldsymbol{r} = \left[ r_x,r_y,r_z \right]^{\mathrm{T}}}$}
satisfies 
${\mathrm{Cov}(\hat{\boldsymbol{r}}) = \mathrm{E} [(\hat{\boldsymbol{r}}-\boldsymbol{r})(\hat{\boldsymbol{r}}-\boldsymbol{r})^{\mathrm{T}} ] \leq \mathbf{F}(\boldsymbol{r})^{-1}}$,
where 
{\color{black}${\mathbf{F}(\boldsymbol{r})\in \mathbb{R}^{3\times 3}}$}
denotes the \ac{FIM}.
Given the Gaussian measurement model~(\ref{eq:measurement_model}) with noise covariance $\mathbf{R}$~(\ref{eq:noise_cov}), we can express the \ac{FIM} as
\begin{equation}
\label{eq:fim}
\mathbf{F}(\boldsymbol{r}) = \mathbf{J}(\boldsymbol{r})^{\mathrm{T}} \mathbf{R}^{-1} \mathbf{J}(\boldsymbol{r}),
\end{equation}
where 
{\color{black}$\mathbf{J}(\boldsymbol{r})\in \mathbb{R}^{|\mathcal{X}^*|\times 3}$}
 denotes the Jacobian matrix of the measurement model with respect to the user location~\cite{rampa2015}. 
Using the differential operator
{\color{black}$\nabla_{\boldsymbol{r}} =[ \frac{\partial}{\partial r_x}, \frac{\partial}{\partial r_y}, \frac{\partial}{\partial r_z} ]^{\mathrm{T}}$,}
the elements of the Jacobian matrix which correspond to $\xi_{l,n}$ are calculated as
\begin{equation}
\begin{split}
\left[ \mathbf{J}(\boldsymbol{r}) \right]_{\xi_{l,n}}
	&= \nabla_{\boldsymbol{r}} h(\boldsymbol{r},\xi_{l,n}) \\
	&= \frac{\phi_{l,n}}{\kappa_{l,n}}  \sum_{u=0}^{N_{\xi_{l,n}}} e^{- \delta_{l,n}^{u}(\boldsymbol{r})/\kappa_{l,n}} 
\nabla_{\boldsymbol{r}} \delta_{l,n}^{u}(\boldsymbol{r}),
\end{split}
\end{equation}
with
\begin{equation}
\nabla_{\boldsymbol{r}} \delta_{l,n}^{u}(\boldsymbol{r}) = \frac{\boldsymbol{r}_{\mathrm{VT}_{l,n}}^{(u)} - \boldsymbol{r}}{\Vert \boldsymbol{r}_{\mathrm{VT}_{l,n}}^{(u)} - \boldsymbol{r} \Vert} + \frac{\boldsymbol{r}_{\mathrm{VR}_{l,n}}^{(N_{\xi_{l,n}}-u)} - \boldsymbol{r}}{\Vert \boldsymbol{r}_{\mathrm{VR}_{l,n}}^{(N_{\xi_{l,n}}-u)} - \boldsymbol{r} \Vert}.
\end{equation}

For the evaluation of the localization accuracy, we can use the \ac{CRLB} to determine the \ac{RMSE} of the location estimate.
Based on the diagonal elements of the \ac{CRLB}, the \ac{RMSE} is lower bounded by
\begin{equation}
\label{eq:rmse}
\mathrm{RMSE} = \sqrt{\mathrm{E}\left[\Vert(\hat{\boldsymbol{r}}-\boldsymbol{r})\Vert^2\right]} \leq 
				\sqrt{\tr\left(\mathbf{F}(\boldsymbol{r})^{-1}\right)}.
\end{equation}

\section{Numerical Results}
Finally, we evaluate the performance of the proposed \ac{MDFL} approach numerically.
Therefore, we assume a fully meshed network of 
${N_{\mathrm{Tx}} = N_{\mathrm{Rx}}=\SI{20}{}}$
circularly arranged, collocated transmitting and receiving nodes.
Each node has a distance of \SI{4}{m} to the network center, as shown in Fig.~\ref{fig:crlb}.
{\color{black}
All nodes are located at the same height and correspond to the height of the body center of the user.
Thus, both the network nodes and the evaluated user locations are placed on the same plane.
}
The multipath propagation environment is characterized by four reflecting surfaces confining an area of 
$\SI{23}{m}\times \SI{15.5}{m}$.
The exact arrangement of the surfaces is shown to scale in Fig.~\ref{fig:crlb_mpc}.
We evaluate both state-of-the-art \ac{DFL} and \ac{MDFL} based on the introduced network.
For illustrating the performance enhancement of \ac{DFL} through \acp{MPC}, we only consider first-order reflections for the \ac{MDFL} system.
{\color{black}
Note that reflections from ground and ceiling are not taken into account in the numerical evaluation.
However, it can be assumed that \acp{MPC} due to these reflections would further improve the performance of \ac{DFL}.
}
%
%
{\color{black}
For the numerical analysis we have refrained from an explicit simulation of the signal propagation, instead we assume perfect association of all considered \acp{MPC}.
This implies that all components of the received signals within the network are assumed to be perfectly estimated and then correctly associated with the respective reflecting sequences~(cf. Sec.~\ref{sec:mdfl_ini}).
}
For both \ac{DFL} and \ac{MDFL}, the parameter set of the measurement model is assumed as ${\phi_{l,n} = \SI{-2.5}{\dB}}$, ${\kappa_{l,n} = \SI{0.05}{m}}$, and ${\sigma_{l,n} = \SI{1.5}{\dB}}$,~$\forall l,n$~\cite{kaltiokallio2017_arti}.
Using (\ref{eq:rmse}), the localization accuracy is evaluated in terms of \ac{RMSE}.
For assessing the spatial localization capabilities, we define the area in which a localization approach achieves an ${\text{RMSE}<\SI{1}{m}}$ as effective observation area.

For the considered environment, Figs.~\ref{fig:crlb_los}~and~\ref{fig:crlb_mpc} provide the resulting \ac{RMSE} values for \ac{DFL} and \ac{MDFL}, respectively.
It can be seen that both approaches cover the area inside the network with high localization accuracy.
For \ac{DFL}, however, this area coincides with the total effective observation area.
In contrast, the effective observation area of \ac{MDFL} spans over the entire environment.
Depending on the topology, \ac{MDFL} covers the areas between network and reflecting surfaces.

\begin{figure}[t!]
	\centering
	\ifthenelse {\boolean{DoubleColumn}}%
	{%
	\includegraphics[width=0.9\columnwidth]{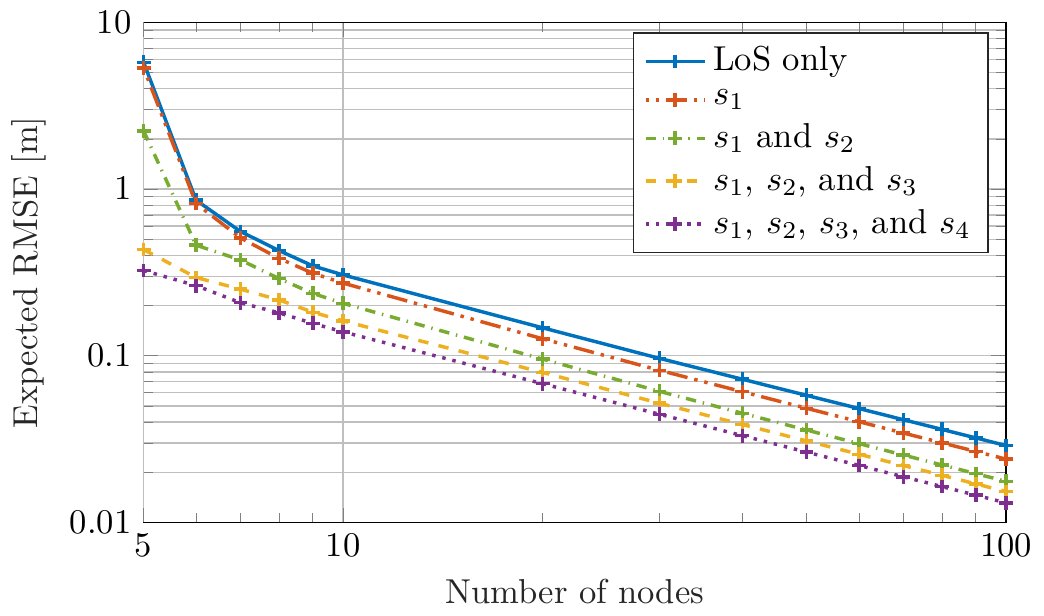}%
	}%
	{%
	\includegraphics[width=0.5\columnwidth]{figures/rmse_numDevices/rmse_numDevices.pdf}%
	}%
	\caption{ \footnotesize Expected RMSE as a function of network nodes, calculated for an area of $\SI{2}{\m} \times \SI{2}{\m}$ in the network center.
	Reflecting surfaces ${\mathcal{S}=\{s_1,s_2,s_3,s_4\}}$ refer to the multipath propagation environment of Fig.~\ref{fig:crlb_mpc}.
	}
\label{fig:crlb_numDevices}
\end{figure}

Apart from a larger coverage, the results in Figs.~\ref{fig:crlb_los}~and~\ref{fig:crlb_mpc} indicate that the localization accuracy has improved for \ac{MDFL} compared to \ac{DFL}.
To further investigate this performance improvement, we calculate an expected \ac{RMSE} for an area of
$\SI{2}{m}\times \SI{2}{m}$
located in the network center, as marked in Fig.~\ref{fig:crlb}.
The expected \ac{RMSE} is determined for different numbers of network nodes and a varying environment.
As before, fully meshed networks of circularly arranged transceiving nodes are assumed and only first-order reflections are considered.
The resulting values are shown in Fig.~\ref{fig:crlb_numDevices}.
For each environment, the expected \ac{RMSE} monotonically decreases with the number of nodes.
Overall, the localization accuracy improves with an increasing number of considered reflection surfaces, i.e., \acp{MPC}.
That means, rich multipath environments are beneficial for \ac{MDFL}.
Thereby, the highest accuracy gain can be achieved for networks employing only a few nodes.
Sparse networks benefit particularly from multipath propagation, since \acp{MPC} compensate for the lack of measurements.


\balance

\acresetall
\section{Conclusion} 
In this letter, we have introduced a \acl{MDFL} approach.
For this novel approach, we have provided a geometrical model describing the physical propagation of \aclp{MPC} and the corresponding signal processing.
Based on the underlying measurement model, we have derived the theoretical performance bound on the localization error allowing to evaluate the proposed approach numerically.
The results show that the identification and consideration of \aclp{MPC} overcomes the constrained spatial localization capabilities of state-of-the-art \acl{DFL} approaches and extends the effective observation area significantly.
Moreover, the localization accuracy is shown to improve with increasing number of \aclp{MPC}.
Thereby, the highest accuracy gain is achieved for sparse networks.

\bibliographystyle{IEEEtran}
\bibliography{bib_v1}

\begin{thebibliography}{10}
\providecommand{\url}[1]{#1}
\csname url@samestyle\endcsname
\providecommand{\newblock}{\relax}
\providecommand{\bibinfo}[2]{#2}
\providecommand{\BIBentrySTDinterwordspacing}{\spaceskip=0pt\relax}
\providecommand{\BIBentryALTinterwordstretchfactor}{4}
\providecommand{\BIBentryALTinterwordspacing}{\spaceskip=\fontdimen2\font plus
\BIBentryALTinterwordstretchfactor\fontdimen3\font minus
  \fontdimen4\font\relax}
\providecommand{\BIBforeignlanguage}[2]{{%
\expandafter\ifx\csname l@#1\endcsname\relax
\typeout{** WARNING: IEEEtran.bst: No hyphenation pattern has been}%
\typeout{** loaded for the language `#1'. Using the pattern for}%
\typeout{** the default language instead.}%
\else
\language=\csname l@#1\endcsname
\fi
#2}}
\providecommand{\BIBdecl}{\relax}
\BIBdecl

\bibitem{shit2019}
R.~C. {Shit}, S.~{Sharma}, D.~{Puthal}, P.~{James}, B.~{Pradhan},
  A.~v.~{Moorsel}, A.~Y. {Zomaya}, and R.~{Ranjan}, ``{Ubiquitous Localization
  (UbiLoc): A} survey and taxonomy on device free localization for smart
  world,'' \emph{IEEE Commun. Surveys Tuts.}, vol.~21, no.~4, pp. 3532--3564,
  Oct.-Dec. 2019.

\bibitem{patwari2010}
N.~{Patwari} and J.~{Wilson}, ``{RF} sensor networks for device-free
  localization: Measurements, models, and algorithms,'' \emph{Proc. IEEE},
  vol.~98, no.~11, pp. 1961--1973, Nov. 2010.

\bibitem{wilson2010}
J.~{Wilson} and N.~{Patwari}, ``Radio tomographic imaging with wireless
  networks,'' \emph{IEEE Trans. Mobile Comput.}, vol.~9, no.~5, pp. 621--632,
  May 2010.

\bibitem{guo2015}
Y.~{Guo}, K.~{Huang}, N.~{Jiang}, X.~{Guo}, Y.~{Li}, and G.~{Wang}, ``An
  exponential-{R}ayleigh model for {RSS}-based device-free localization and
  tracking,'' \emph{IEEE Trans. Mobile Comput.}, vol.~14, no.~3, pp. 484--494,
  Mar. 2015.

\bibitem{rampa2015}
V.~{Rampa}, S.~{Savazzi}, M.~{Nicoli}, and M.~{D’Amico}, ``Physical modeling
  and performance bounds for device-free localization systems,'' \emph{IEEE
  Signal Process. Lett.}, vol.~22, no.~11, pp. 1864--1868, Nov. 2015.

\bibitem{kaltiokallio2017_arti}
O.~{Kaltiokallio}, R.~{Jäntti}, and N.~{Patwari}, ``{ARTI}: An adaptive radio
  tomographic imaging system,'' \emph{IEEE Trans. Veh. Technol.}, vol.~66,
  no.~8, pp. 7302--7316, Aug. 2017.

\bibitem{beck2016}
B.~{Beck}, X.~{Ma}, and R.~{Baxley}, ``Ultrawideband tomographic imaging in
  uncalibrated networks,'' \emph{IEEE Trans. Wireless Commun.}, vol.~15, no.~9,
  pp. 6474--6486, Sep. 2016.

\bibitem{schmidhammer2020}
M.~Schmidhammer, M.~Walter, C.~Gentner, and S.~Sand, ``Physical modeling for
  device-free localization exploiting multipath propagation of mobile radio
  signals,'' in \emph{Proc. 14th Eur. Conf. on Antennas and Propag. (EuCAP
  2020)}, Apr. 2020.

\bibitem{molisch2009}
A.~F. Molisch, ``Ultra-wide-band propagation channels,'' \emph{Proc. IEEE},
  vol.~97, no.~2, pp. 353--371, Feb. 2009.

\bibitem{meissner2014}
P.~Meissner, E.~Leitinger, and K.~Witrisal, ``{UWB} for robust indoor tracking:
  Weighting of multipath components for efficient estimation,'' \emph{IEEE
  Wireless Commun. Lett.}, vol.~3, no.~5, pp. 501--504, Oct. 2014.

\bibitem{gentner2016}
C.~{Gentner}, T.~{Jost}, W.~{Wang}, S.~{Zhang}, A.~{Dammann}, and U.~{Fiebig},
  ``Multipath assisted positioning with simultaneous localization and
  mapping,'' \emph{IEEE Trans. Wireless Commun.}, vol.~15, no.~9, pp.
  6104--6117, Sep. 2016.

\bibitem{fleury1999}
B.~H. Fleury, M.~Tschudin, R.~Heddergott, D.~Dahlhaus, and K.~I. Pedersen,
  ``{Channel parameter estimation in mobile radio environments using the SAGE
  algorithm},'' \emph{IEEE J. Sel. Areas Commun.}, vol.~17, no.~3, pp.
  434--450, Mar. 1999.

\bibitem{schuhmacher2008}
D.~{Schuhmacher}, B.~{Vo}, and B.~{Vo}, ``A consistent metric for performance
  evaluation of multi-object filters,'' \emph{IEEE Trans. Signal Process.},
  vol.~56, no.~8, pp. 3447--3457, July 2008.

\end{thebibliography}

\end{document}